\documentclass[twocolumn,english,british,aps,prl,floatfix,superscriptaddress,preprintnumbers,showpacs,longbibliography]{revtex4-2}
\usepackage[T1]{fontenc}
\usepackage[utf8]{inputenc}
\usepackage{babel}
\usepackage{verbatim}
\usepackage{amsmath}
\usepackage{amsthm}
\usepackage{amssymb}
\usepackage{graphicx}
\usepackage{geometry}
\usepackage{physics}
\usepackage{dsfont}
\usepackage[pdfusetitle,
 bookmarks=true,bookmarksnumbered=false,bookmarksopen=false,
 breaklinks=false,pdfborder={0 0 1},backref=false,colorlinks=false]
 {hyperref}

\graphicspath{{images}}

\usepackage{xcolor}
\makeatletter

\providecommand{\tabularnewline}{\\}

\theoremstyle{plain}
\newtheorem*{thm*}{\protect\theoremname}
\theoremstyle{plain}
\newtheorem*{cor*}{\protect\corollaryname}

\makeatother

\addto\captionsbritish{\renewcommand{\corollaryname}{Corollary}}
\addto\captionsbritish{\renewcommand{\theoremname}{Theorem}}
\addto\captionsenglish{\renewcommand{\corollaryname}{Corollary}}
\addto\captionsenglish{\renewcommand{\theoremname}{Theorem}}
\providecommand{\corollaryname}{Corollary}
\providecommand{\theoremname}{Theorem}

\newcommand{\prlsection}[1]{\textbf{\textit{#1.}}}

\begin{document}
\title{Quantum-to-Classical Computability Transition via Negative Markov Chains}

\author{Hugo L\'oio}
\affiliation{Laboratoire de Physique Th\'eorique et Mod\'elisation, CNRS UMR 8089, CY Cergy Paris Universit\'e, 95302 Cergy-Pontoise Cedex, France}
\affiliation{JEIP, UAR 3573 CNRS, Collège de France, PSL Research University,
11 Place Marcelin Berthelot, 75321 Paris Cedex 05, France}

\author{Jacopo De Nardis}
\affiliation{Laboratoire de Physique Th\'eorique et Mod\'elisation, CNRS UMR 8089, CY Cergy Paris Universit\'e, 95302 Cergy-Pontoise Cedex, France}
\affiliation{JEIP, UAR 3573 CNRS, Collège de France, PSL Research University,
11 Place Marcelin Berthelot, 75321 Paris Cedex 05, France}

\author{Tony Jin}
\affiliation{Universit\'e C\^ote d'Azur, CNRS, Centrale Med, Institut de Physique de Nice, 06200 Nice, France}

\begin{abstract}
We develop a recently introduced representation of quantum dynamics based on sampling negative Markov chain processes. By introducing particles and antiparticles, this formalism maps generic quantum dynamics onto a Markov process defined over an exponentially large configuration space. Within this framework, quantum complexity arises from the proliferation of stochastic particles, which ultimately renders classical simulation and sampling intractable beyond a certain timescale. In the presence of noise, we demonstrate that for any unitary evolution generated by a linear combination of local or pairwise interactions, there exists at least one noise channel that effectively classicalizes the system by suppressing particle growth and making Monte Carlo sampling efficient. As a corollary, we show that for this class of unitaries, the dynamics of an open quantum spin chain subject to depolarizing noise undergoes an exact transition to classical simulability once the noise strength exceeds a critical threshold which can be efficiently determined for any model.
\end{abstract}

\maketitle

\prlsection{Introduction}
A major promise of quantum simulation is the ability to access systems and dynamical regimes that are beyond the reach of classical computers~\cite{A_Aaronson2011,B_Bremner2016}. 
This is widely expected to hold for sufficiently well-isolated quantum systems, where interactions with the external environment are negligible or very weak~\cite{shor1996fault, kitaev2003faulttolerant, aharonov2008faulttolerant}. Current quantum platforms instead operate in the noisy intermediate-scale quantum (NISQ) regime, where the number of qubits is limited and noise remains significant~\cite{preskill2018quantum}. It is therefore crucial to understand when the simulation of a given quantum system remains genuinely hard for classical computers, and when it can instead be efficiently reproduced classically. This question is directly relevant to experimental claims of quantum advantage~\cite{arute2019quantum,zhong2020quantum,kim2023evidence,morvan2024phase,zhong2021phase,deng2023gaussian,liu2025robust,decross2025computational,king2025beyond}, many of which have subsequently been challenged by increasingly powerful classical simulation methods~\cite{gao2018efficient,huang2020classical,pan2021simulating,pan2022solving,pan2022simulation,liu2021closing,tindall2024efficient,beguvsic2023fast,oh2022classical,oh2024classical,mauron2025challenging,tindall2025dynamics,TWA-marino}. In many such cases, classical algorithms explicitly exploit the presence of noise: noise can drive the output distribution toward a trivial one~\cite{aharonov1996limitations}, or effectively reduce the accessible Hilbert space~\cite{aharonov2023polynomial,schuster2024polynomial}, thereby rendering classical simulation efficient.

In recent years, several works have shown~\cite{aharonov2023polynomial,gao2018efficient,schuster2024polynomial} that efficient classical sampling is possible in generic noisy quantum systems whenever the output distribution satisfies anticoncentration. Moreover, for broad classes of circuits above a constant noise threshold, one generically finds a critical noise rate beyond which efficient classical simulation becomes possible through methods based on percolation arguments~\cite{trivedi2022transitions,rajakumar2025polynomial,nelson2024polynomialtimeclassicalsimulationnoisy}.

In this paper, we uncover a \textit{new noise-induced quantum-to-classical computability transition based on the feasibility of Monte Carlo sampling} and an associated sign-problem transition. We investigate the possibility of representing quantum evolution through Monte Carlo sampling in configuration space. Generically, these dynamics cannot be formulated as a Markov chain because the effective transition rates are not always positive. This difficulty can nevertheless be overcome by introducing two species of stochastic particles and antiparticles that encode the sign structure. In this representation, however, quantum complexity reappears in the exponential growth of the number of particles with time, making the computation infeasible beyond a certain timescale. Our main result is that there always exists a critical amount of noise above which all effective transition rates become positive, thereby restoring a genuine stochastic description and making Monte Carlo sampling efficient.

 \textit{The transition we observe appears to be distinct from those previously studied }\cite{trivedi2022transitions,rajakumar2025polynomial,nelson2024polynomialtimeclassicalsimulationnoisy} and is not directly tied to standard complexity quantifiers such as entanglement. Instead, it is governed solely by a transition between a simulable regime with only positive weights and one with negative weights (similarly to \cite{signTensorN}). In the latter, the corresponding particle proliferation indicates the severity of the quantum mechanical sign problem \cite{Loh1990,Troyer2005}.
 Remarkably, we find the existence of a \textit{gauge freedom} in the Markov chain description of quantum dynamics that plays a role analogous to the basis rotations used to tame the sign problem in equilibrium quantum Monte Carlo \cite{Chandrasekharan1999,Li2015,Marvian2019}. Finally, beyond its conceptual interest, our algorithm provides a quantitative method to determine, for a generic quantum spin chain, \textit{the critical noise rate} at which this transition occurs.

\begin{figure}
\centering{}\includegraphics[width=1\columnwidth]{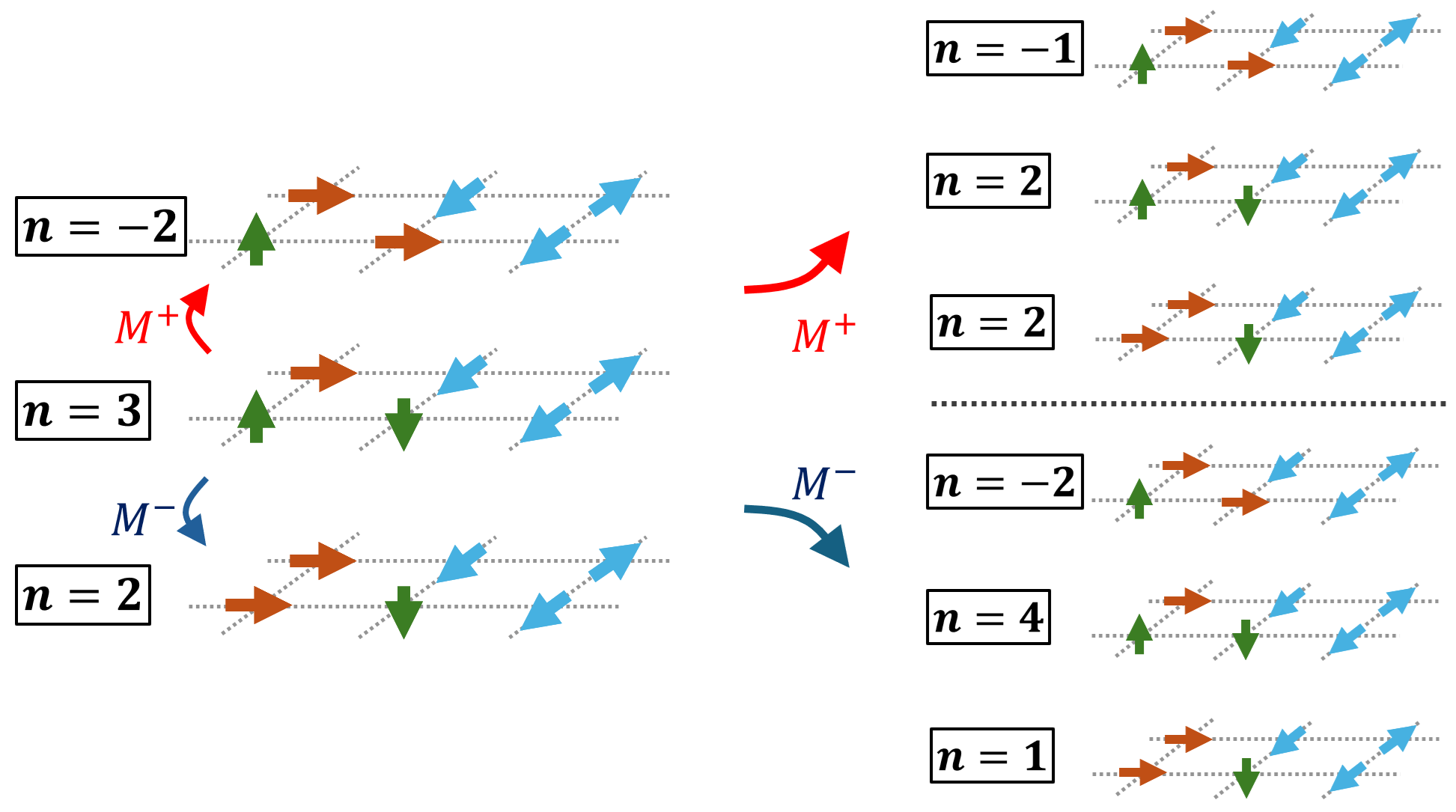}
\caption{Schematic depiction of the transition rules. The negative Markov chain
formalism describes a process that takes place on all possible $6^{N}$
configurations. Each configuration is weighted by an occupation number
that can be negative. The Hamiltonian determines the transition rates
$M^{+}$ and $M^{-}$ between different configurations. A transition
with $M_{CC'}^{+}$ decreases the occupation number of the source configuration
$n_{C'}$ by $1$ while increasing the target occupation number $n_{C}$ by $1$.
The transition $M_{CC'}^{-}$ does \emph{the reverse}, i.e. it increases
$n_{C'}$ by $1$ and decreases $n_{C}$ by $1$. The computational complexity arises from the need to track the occupation number of every copy. The classical CTMC
case is recovered when $M^{-}=0$. In this case, only one configuration at a time is occupied and the dynamics can be simulated via classical Monte Carlo.}
\label{fig:Picture_replicas}
\end{figure}

We first introduce the formalism of negative Markov chains (NMCs) and establish an exact mapping of quantum dynamics onto a classical stochastic process in an exponentially large configuration space. We then show that this classical representation possesses an intrinsic gauge freedom, which can be exploited to enforce complete cancellation of particle proliferation once the noise exceeds a critical threshold. Finally, we illustrate our approach with an application to the noisy transverse-field Ising model (TFIM).

\prlsection{Negative Markov chains}
We begin by reviewing the NMC formalism for quantum spin chains, as introduced
in \cite{Negativerates_Vollering,jin2025classicalrep}. Consider a quantum spin chain in $d$ dimensions, of length $L$ and total site number $N=L^d$, evolving under
the Lindbladian ${\cal L}$ \cite{F_Lindblad1976,G_Gorini1976}: 
\begin{equation}
d_{t}\rho={\cal L}\left(\rho\right):=-i\left[H,\rho\right]+\sum_{k}A_{k}^{\dagger}\rho A_{k}-\frac{1}{2}\left\{ A_{k}A_{k}^{\dagger},\rho\right\} .\label{eq: quantum-dyn}
\end{equation}
Our goal is to describe this time evolution in terms of a Markov chain
on a classical configuration space ${\cal C}$ made of the $6^{N}$
elements 
\begin{equation} \label{eq:classicalconfigurationspace}
{\cal C}=\left(\left\{ +,-\right\} \times\left\{ x,y,z\right\} \right)^{N}  \, ,
\end{equation}
where $\left\{ +,-\right\}$ denotes the orientation of the spin
along a given axis. An element $C\in{\cal C}$ is a \emph{classical
configuration} of the spin chain. To each $C$, we can naturally associate
the Hilbert space vector $\left|C\right\rangle$. Note that $\left\{ \left|C\right\rangle \right\} _{C\in{\cal C}}$
is an overcomplete basis, as it spans all three axes for each spin.
Let $\mathbb{P}_{C}:=\left|C\right\rangle \left\langle C\right|$
and $p_{C}\left(t\right)$ be the probability of finding the system
in the classical configuration $C$ at time $t$: 
\begin{equation}
p_{C}(t):=\frac{1}{3^{N}}{\rm tr}\left(\rho_{t}\mathbb{P}_{C}\right).
\end{equation}
Since $p_{C}\geq0$ and $\sum_{C\in{\cal C}}p_{C}=1$, $p_{C}$ is
a well-defined probability distribution. Its time evolution is inherited
from Eq.~(\ref{eq: quantum-dyn}):
\begin{equation}
d_{t}p_{C}=\frac{1}{3^{N}}{\rm tr}\left(\rho_{t}{\cal L}^{*}\left(\mathbb{P}_{C}\right)\right),
\end{equation}
with ${\cal L}^{*}$ the dual of ${\cal L}$. Since ${\cal L}^{*}\left(\mathbb{P}_{C}\right)$
is Hermitian (but not necessarily traceless), it can in general be
decomposed on the basis of Pauli strings (including the identity). Since these
Pauli strings can in turn be decomposed on the projectors $\mathbb{P}_{C}$,
we end up with
\begin{equation}
{\cal L}^{*}\left(\mathbb{P}_{C}\right)=\sum_{C'}M_{CC'}\mathbb{P}_{C'}.\label{eq:L_decomposition}
\end{equation}
In general, this decomposition is not unique due to the identity operator
in a Pauli string: its expansion into projectors is inherently ambiguous---a
crucial gauge freedom that we explore later. Projecting onto $\rho$ one
obtains
\begin{equation}
d_{t}p_{C}=\sum_{C'\neq C}M_{CC'}p_{C'}-M_{C'C}p_{C},\label{eq:CTMC_negative}
\end{equation}
where we used the relation $M_{CC}=-\sum_{C'\neq C}M_{C'C}$, which
follows from probability conservation. The previous equation is nearly
that of a continuous-time Markov chain (CTMC) \footnote{Note that in our convention, $M_{CC'}$ means a transition from state
$C'$ to $C$.}. The coefficients $M_{CC'}$
are real due to the Hermiticity of ${\cal L}^{*}\left(\mathbb{P}_{C}\right)$,
but they do not satisfy the required positivity. This places us in the realm
of CTMCs with negative rates, a scenario only recently examined in \cite{Negativerates_Vollering,jin2025classicalrep}. 
First, we define the \emph{entirely positive} transition rates for $C\neq C'$,
\[
M_{CC'}^{\pm}:=\pm\theta\left(\pm M_{CC'}\right)M_{CC'},
\]
where $\theta$ is the Heaviside function. We then double the configuration
space and denote \emph{particle} states with a $\bullet$ and \emph{antiparticle}
states with a $\circ$: ${\cal C}\to{\cal C}\times\left\{ \bullet,\circ\right\}.$
The probability $p_{C}$ of the original process is decomposed as
the difference
\begin{equation}
p_{C}=p_{C}^{\bullet}-p_{C}^{\circ}.\label{eq:Annihilation}
\end{equation}
After a few elementary manipulations, one sees that Eq.~(\ref{eq:CTMC_negative})
can be obtained from
\begin{equation}\label{eq:probabilitesparticle/antiparticle}
\begin{aligned}
d_{t}p_{C}^{\bullet/\circ}= & \sum_{C'\neq C}\big(M_{CC'}^{+}p_{C'}^{\bullet/\circ}-M_{C'C}^{+}p_{C}^{\bullet/\circ}\\
 & +M_{CC'}^{-}p_{C'}^{\circ/\bullet}-M_{C'C}^{-}p_{C}^{\bullet/\circ}\big)+V_{C}p_{C}^{\bullet/\circ}
\end{aligned}
 \, ,
\end{equation}
with $V_{C}:=2\sum_{C'\neq C}M_{C'C}^{-}$. The previous equation
now defines a well-defined Markov process with entirely positive
rates given by $M^{+}$ and $M^{-}$ and \emph{creation} or \emph{branching}
rates given by $V_{C}$ \cite{H_Harris1963}. Due to the latter, the state of the system for a given realization
of the process is now described by a collection of $2\times6^{N}$
numbers $n_{C}^{\bullet/\circ}$ indicating the occupation of configuration
$C$ in terms of particles and antiparticles.
As allowed by Eq.~(\ref{eq:Annihilation}),
we impose that whenever a particle and an antiparticle meet, they annihilate
each other. Hence, a single number $n_{C}:=n_{C}^{\bullet}-n_{C}^{\circ}\in\mathbb{Z}$
suffices to keep track of the state of the system. The configuration
space is then $\mathbb{Z}^{\cal C}$, which means that we
have $\Omega$ replicas of the original spin chain, where $\Omega=\sum_{C\in{\cal C}}\left|n_{C}\right|$.
A \emph{state} $\eta\in\mathbb{Z}^{{\cal C}}$ is the collection of
numbers $\left\{ n_{C}\right\} _{C\in{\cal C}}$. Let $s_{C}:={\rm sign}\left(n_{C}\right)$.
The update rules can be read from Eq.~(\ref{eq:probabilitesparticle/antiparticle})
to be:
\begin{equation}
\eta\to\eta\pm s_{C'}\left(-\delta_{C'}+\delta_{C}\right)\text{ with probability \ensuremath{\left|n_{C'}\right|M_{CC'}^{\pm}dt}},\label{eq:transition_rules}
\end{equation}
and are illustrated schematically in Fig.~\ref{fig:Picture_replicas}.
The $M^{+}$ transition has the usual interpretation for a CTMC, where a state with configuration $C'$ goes to $C$, while $M^{-}$ can be interpreted as a transition from $C'$ to $C$ carrying a \emph{negative sign}, see Fig.~\ref{fig:Picture_replicas}. Starting with $\Omega(t=0)=1$, the original probabilities are extracted by averaging over different
realizations of the process, i.e.
\begin{equation}
p_{C}=\mathbb{E}\left[n_{C}\right].
\end{equation}
Remark that while $p_{C}$ is constrained to be positive, for a given
realization, $n_{C}$ can be `off-shell', i.e., it may take any integer
value, positive or negative. Finally, we note that the rules in Eq.~(\ref{eq:transition_rules})
preserve the total relative number of particles $\sum_{C}n_{C}$.

A key advantage of obtaining a CTMC description of the quantum dynamics is that it enables the simulation
of individual trajectories, paving the way for classical Monte Carlo
methods \cite{I_Gillespie1977}. In a given realization, the simulation proceeds by updating
the occupation numbers $\eta$ according to the transition rules in
Eq.~(\ref{eq:transition_rules}). Let $\Omega_{\text{occ}}=\sum_{C}\left(1-\delta_{n_{C},0}\right)$
denote the number of occupied configurations; the size of $\eta$
is therefore $\Omega_{\text{occ}}$ and is time dependent. By
organizing $\eta$ in a tree structure, the cost per fixed time step scales as
\begin{equation}
{\cal O}\left(\Omega\log \Omega_{\rm occ}\log N\right),
\end{equation}
where $\log\Omega_{{\rm occ}}$ is the cost of searching for an
element in $\eta$, $\Omega$ is the number of times this operation
must be performed, and $\log N$ is the computational complexity required
to compute a transition for a single particle (see App.~\ref{App:Numerics} for more details).
A simple heuristic picture for the growth of $\Omega\left(t\right)$
is as follows. At early times, when the number of occupied configurations is low, we expect the growth to be controlled by the negative weights $V_{C}\sim N$ and thus to be exponential in time and system size,
\begin{equation}
\Omega\left(t\right)\sim e^{\mu tN}.
\end{equation}
Assuming that the configurations are populated uniformly,
a simple estimate for $\mu$ is:
\begin{equation}\label{eq_mu_prediction}
\mu:=\frac{2d}{6^{k}}\sum_{C\neq C'}M_{CC'}^{\left(k\right)-},
\end{equation}
where $d$ is the dimension and $M^{\left(k\right)-}$ is the local negative transition matrix acting on $k$ sites.
At late times, the growth is damped by the fact that particles and
antiparticles start to annihilate.
If saturation occurs, and again assuming uniform population, then the saturation value is (see App.~\ref{App:Predictions} for
details):
\begin{equation}
\Omega_{{\rm sat}}:=-2\left(6^{N}\right)\log\left(1-\frac{\sum_{C\neq C'}M_{CC'}^{\left(k\right)-}}{\sum_{C\neq C'}\left| M_{CC'}^{\left(k\right)} \right|}
\right).
\end{equation}
While the late-time behavior is more favorable, the early exponential
growth in $t$ and $N$ prevents the simulation of large system sizes.
Nevertheless, in the following, we demonstrate that, for open quantum systems, the
growth of $\Omega$ can be drastically reduced---and in some cases
eliminated entirely---thereby reducing the simulation to a simple CTMC with
a per-step complexity of $\mathcal{O}(\log N)$.

\prlsection{Markov matrices and gauge freedom} We now need to obtain the Markov matrix from a given Lindbladian ${\cal L}$,
i.e. we need to invert the relation (\ref{eq:L_decomposition}). Consider
in full generality a local dynamics ${\cal L}=\sum_{j}{\cal L}_{j}^{k}$,
where ${\cal L}_{j}^{k}$ acts on $k$ qudits. Let $|m)$ be a canonical basis of local operators of dimension $d^{2k}$, $\left\{ |m=(a-1)d^{k}+b)\right\} {}_{m=1}^{d^{2k}}:=\left\{ \left|a\right\rangle \left\langle b\right|\right\} _{a,b=1}^{d^{k}}$,
and let $q\geq d^{2}$ be the number of local configurations for the
classical space (for spins, $q=6$).
We introduce the decomposition $\mathbb{P}_{C}=:\sum_{m}A_{C,m}|m)$,
where $A$ is a $q^{k}\times d^{2k}$ matrix, i.e., the rows of $A$ correspond to the vectorized $\mathbb{P}_C$.
We impose that $\sum_C \mathbb{P}_C=\mathbb{I}$ to enforce conservation of probability \footnote{Let $\vec{1}$ be the $q^{k}$-dimensional vector with all entries equal to one. The conservation of probability for the matrix $M$ translates
into $\vec{1}^{T}M=0$, which is equivalent to ${\cal L}^{*}\left(\sum_{C}\mathbb{P}_{C}\right)=0$.
A sufficient condition to enforce this relation is then $\sum_{C}\mathbb{P}_{C}\propto\mathbb{I}$.
}. Additionally, the configurations
$C$ are chosen such that a given $|m)$ always admits a decomposition
in terms of $\mathbb{P}_{C}$ \footnote{Note that the classical spin configuration basis $\mathcal{C}=\{+,-\} \times \{ x,y,z \}$ is the smallest basis fulfilling the two previous criteria.}. Hence, $A^{T}$ is a \emph{surjective}
map, $A$ is \emph{injective}, and admits a Moore-Penrose left inverse
denoted $A^{+}$, with $A^{+}:=\left(A^{\dagger}A\right)^{-1}A^{\dagger}$
and $A^{+}A=\mathbb{I}$. Now let ${\cal {\cal M}}$ be the matrix
representation of ${\cal L^*}$ in the flattened basis $|m)$: ${\cal L^*}\left(|n)\right)=:\sum_{m}{\cal M}_{n,m}|m)$,
which is always unique. Eq.~(\ref{eq:L_decomposition}) can be recast
as
\begin{align}
A{\cal M} & =MA.\label{eq:AM=MA}
\end{align}
Its non-unique solution is
\begin{equation}
M=A\mathcal{M}A^{+} \, .
\end{equation}
In this form, the gauge transformation mentioned earlier becomes apparent. Indeed,
let $\Lambda$ be a $q^{k}\times q^{k}$ matrix such that the rows
of $\Lambda$ belong to the null space of $A^{T}$. Then $\Lambda A=0$
and $M+\Lambda$ is also a solution of \eqref{eq:AM=MA}.
Since each row of $\Lambda$ belongs to the null space of $A^{T}$
and the latter is surjective, they can be parametrized by $q^{k}-d^{2k}$
elements. Multiplying this by the number of rows, which is $q^{k}$,
and adding the constraint of probability conservation $\sum_{C}\Lambda_{C,C'}=0$,
we end up with $\left(q^{k}-d^{2k}\right)\left(q^{k}-1\right)$ independent
elements parametrizing the gauge degrees of freedom. For the two
cases of primary interest to us, $d=2$ and $k=1$ or $k=2$, this
results in 10 and 700 degrees of freedom, respectively. In the following,
we exploit these gauge degrees of freedom to optimize the cancellation
between the positive terms arising from the open-system dynamics and
the negative contributions from the unitary evolution.

We now provide explicit expressions for the Markov matrices. For
the unitary part of the evolution, we consider any local or pairwise
Pauli strings
\begin{equation}
H=\sum_{j,a}h_{j}^{a}\sigma_{a}^{j}+\sum_{j\neq k,a,b}J_{a,b}^{j,k}\sigma_{a}^{j}\sigma_{b}^{k},\label{eq:Hamiltonian}
\end{equation}
where the coefficients $h$ and $J$ are arbitrary and the indices
$a$, $b$ denote the directions $\left\{ x,y,z\right\}$. Similarly,
for open systems, we consider
\begin{equation}
{\cal L}=\sum_{j,a}\nu_{j}^{a}{\cal L}_j^a+\sum_{j\neq k,a,b}\mu_{j,k}^{a,b}{\cal L}^{a,b}_{j,k},\label{eq:Lindblad}
\end{equation}
where we defined $\mathcal L ^a:=\mathcal {L}\left[\sigma_a\right]$, $\mathcal L ^{a,b}:=\mathcal {L}\left[\sigma_a \sigma_b\right]$ with ${\cal L}\left[A\right]\left(\rho\right)=A\rho A^{\dagger}-\frac{1}{2}\left\{ A^{\dagger}A,\rho\right\}$.
An explicit computation provides, up to the arbitrary gauge choice,
the corresponding Markov matrix elements, reported in Table.~\ref{tab:Summary-of-the-terms}.

\begin{table}
\begin{centering}
\begin{tabular}{|c|c|}
\hline
Hamiltonian/Lindblad & Markov matrix elements $M_{C,C'}$\tabularnewline
\hline
\hline
$H=\sigma_{a}$ & $-ss'\varepsilon_{a,\alpha,\alpha'}$\tabularnewline
\hline
$H=\sigma_{a}^{j}\sigma_{b}^{k}$ & $-\frac{1}{2}\big(\varepsilon_{a,\alpha,\alpha'}s_{1}s'_{1}\left(s'_{2}+s_{2}\delta_{b,\beta}\right)\delta_{b,\beta'}$\tabularnewline
 & $+(s_{1},\alpha)\leftrightarrow(s_{2},\beta)\big)$\tabularnewline
\hline
${\cal L}^{*}\left[\sigma_{a}\right]$ & $\delta_{(S_{a,\alpha}s,\alpha),C'}-\delta_{C,C'}$\tabularnewline
\hline
${\cal L}^{*}\left[\sigma^{\pm}\right]$ & $-\frac{s}{4}\delta_{\alpha,\alpha'}\left(s'\bar{\delta}_{z,\alpha}+2\delta_{z,\alpha}\left(s'\mp1\right)\right)$\tabularnewline
\hline
${\cal L}^{*}\left[\sigma_{a}^{j}\sigma_{b}^{k}\right]$ & $\delta_{\left(S_{a,\alpha}s_{1},\alpha,S_{b,\beta}s_{2},\beta\right),C'}-\delta_{C,C'}$\tabularnewline
\hline
\end{tabular}
\par\end{centering}
\caption{Summary of the different terms considered in this study, together with the
expression of their action on projectors on the spin eigenbasis and
the corresponding Markov matrix elements. Our notation conventions
for the indices are $C:=\left(s,\alpha\right)$ for a single site
and $C:=\left(s_{1},\alpha,s_{2},\beta\right)$ for two sites, where
in each case the first index denotes the axis and the second the
sign of the spin. We also introduced the notation $\bar{\delta}_{a,\alpha}:=1-\delta_{a,\alpha}$
and $S_{a,\alpha}=\delta_{a,\alpha}-\bar{\delta}_{a,\alpha}$.}
\label{tab:Summary-of-the-terms}
\end{table}

First, we remark that all of these expressions are such that each row
and column of $M$ sums to $0$. The former condition is necessary
to ensure probability conservation, while the latter implies that
the uniform probability distribution is a stationary solution. From
the explicit expressions of the unitary terms, we see that there are
as many negative terms as positive ones. Hence, one generically
expects unbounded growth of particle numbers. On the other hand, the
open-system terms have only negative contributions on the diagonal,
giving them a proper interpretation as classical CTMC rates. Importantly, in
these terms only the polarization sign index can be off diagonal,
which gives 3 possible combinations:
\begin{equation}
\{\delta_{s_{1},s'_{1}}\delta_{-s_{2},s'_{2}},\delta_{-s_{1},s'_{1}}\delta_{s_{2},s'_{2}},\delta_{-s_{1},s'_{1}}\delta_{-s_{2},s'_{2}}\}.
\end{equation}
Combined with the $9$ possibilities for fixing the polarization axis
for both spins, this leads to $27$ possibilities for the positive
off-diagonal terms of the open-system contribution. For later convenience we denote
this set by $\mathfrak{L}$.

\prlsection{Classical computability}
We are now in a position to state our main result: 
\begin{thm*} \emph{Classical computability of open quantum systems.} Given a Hamiltonian
$H$ of the form (\ref{eq:Hamiltonian}), there always exists a combination
of noise terms ${\cal L}$ of the form (\ref{eq:Lindblad}) with finite weights and a gauge
transformation $\Lambda$ such that the associated CTMC has only purely
positive transition rates and is thus computable with a classical Monte
Carlo algorithm with complexity ${\cal O}\left(\log N\right)$ for
a single time step, where $N$ is the size of the system.
\end{thm*}
With the following corollary:
\begin{cor*}
{\emph{Universal noise.}} Given a Hamiltonian $H$ of the form (\ref{eq:Hamiltonian})
and the linear two-spin depolarizing noise $\gamma\left(\sum_{a,j}{\cal L}^{a}_j+\sum_{a,b,\langle i,j\rangle}{\cal L}^{a,b}_{i,j}\right)$,
there exists a gauge $\Lambda$ such that the associated CTMC becomes
classical when $\gamma$ exceeds a finite critical value
$\gamma_{c}$, whose value is model dependent.
\end{cor*}
The detailed proof is given in App.~\ref{sec:Proof}; here we only
provide a sketch. As explained above, the Markov elements of the dissipative terms all belong to $\mathfrak{L}$. It turns out that we can systematically cast the negative weights of the single-site
and pairwise Hamiltonian terms in the same form by making use of
the respective gauge transformations:
\begin{equation}\label{eq:canonical_gauge}
\begin{aligned}
\Lambda_{CC'}^{a}:= & \varepsilon_{a,\alpha,\alpha'}^{2}-\bar{\delta}_{a,\alpha}\delta_{\alpha,\alpha'}\\
\Lambda_{CC'}^{a,b}:= & \frac{1}{2}\big(\left(\varepsilon_{a,\alpha,\alpha'}^{2}-\bar{\delta}_{a,\alpha}\delta_{\alpha,\alpha'}\right)\delta_{b,\beta'}\left(1+\delta_{\beta,\beta'}s_{2}s'_{2}\right) \\
 & +\bar{\delta}_{a,\alpha}\delta_{\alpha,\alpha'}\left(\delta_{b,\beta}\left(1-3\delta_{b,\beta'}\right)+\delta_{b,\beta'}-\delta_{\beta,\beta'}\right) \\
 & +(s_{1},\alpha)\leftrightarrow(s_{2},\beta)\big) \, ,
\end{aligned}
\end{equation}
where $C$ denotes the single-site index $(\alpha,s)$ and, for two sites,
$C:=\left(\alpha,s_{1},\beta,s_{2}\right)$. Since all the possible
terms in $\mathfrak{L}$ can be obtained from the set of Lindblad
jump operators we consider, this proves the theorem. The proof
of the corollary follows from the fact that the uniform two-spin depolarizing noise contains all the terms in $\mathfrak{L}$. Thus, there is necessarily
a critical value $\gamma_{c}$ above which all weights become
positive.

\begin{figure}[t!]
    \centering
    \includegraphics[trim={2 5 2 11}, clip, width=\linewidth]{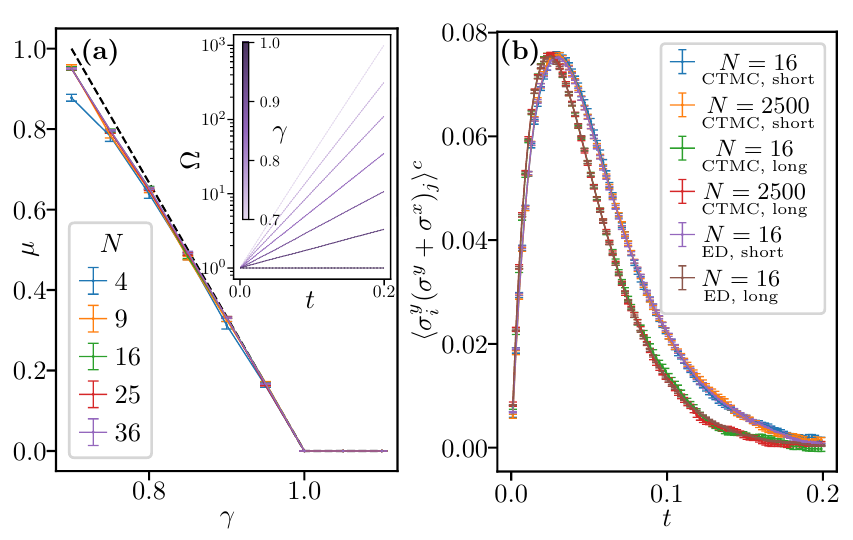}
    \caption{
    Numerical results for the 2D TFIM ($J = 1/2, h = 1$)  with noise given by \eqref{eq:TFIM_noise} with a prefactor $\gamma$ (yielding classical dynamics for $\gamma \geq 1$). 
    The initial state is given by \eqref{eq_bell_pairs_state_main}, with pairs randomly chosen and fixed for all samples, except for those specifically selected for the curves in (b).
    (a) Particle growth rate as a function of $\gamma$ for different qubit numbers $N$.
    Dashed lines show the prediction of \eqref{eq_mu_prediction}.
    (a, inset) Particle number $\Omega$ as a function of time $t$, for different $\gamma$ and $N = 36$.
    (b) Two-point connected correlation $\langle \sigma^y_i (\sigma^y + \sigma^x)_j\rangle^c$ as a function of time $t$, for different qubit numbers $N$, comparing CTMC and ED simulations.
    The sites $i,j$ are maximally entangled in a Bell pair at $t = 0$.
    Short correlations correspond to neighboring $i,j$, and long correlations correspond to $i,j$ being as far apart as possible.
    CTMC averaging was performed over $10^4$ and $10^8$ trajectories for (a) and (b), respectively.
    The ED simulations correspond to pure-state-averaged trajectories following a Lindbladian unraveling, over $10^6$ trajectories.
    }
    \label{fig_tfim}
\end{figure}

As explained above, the noise that makes the CTMC classical is in general non-unique.
In order to find a suitable set of noise terms, one
must solve a set of linear coupled inequalities of the form
\begin{equation}
Q.\vec{x}\geq\vec{y}\left[H\right],
\end{equation}
where $Q$ is a $27\times15$ matrix whose columns give the decomposition
of the Lindblad terms in (\ref{eq:Lindblad}) in terms
of elements of $\mathfrak{L}$, $\vec{x}$ is a $15$-dimensional vector
containing the weights $\nu$ and $\mu$, and $\vec{y}$ is a $27$-dimensional
vector containing the negative terms of $H$ written in the gauge (\ref{eq:canonical_gauge}).
Thus, finding an optimal noise reduces to finding the best $\vec{x}$
for the above inequality while optimizing a given cost function, for
instance the ${\rm L}^{1}$ norm of $\vec{x}$, which is a straightforward
numerical task. For example, in dimension $2$, the transverse-field
Ising model (TFIM)
\begin{equation}\label{eq:TFIM_Hamiltonian}
H=\sum_{j}h\sigma_{z}^{j}+\sum_{\langle i,j\rangle}J\sigma_{x}^{i}\sigma_{x}^{j} \, ,
\end{equation}
with $\langle i,j\rangle$ neighbouring sites, 
is made classical by adding the following Lindblad terms on each link:
\begin{equation}
\gamma\left|J\right|({\cal L}^{x,x}+\sum_{a,b}\frac{{\cal L}^{a,b}}{2})+
\frac{\gamma \left|h\right|}{4}\left({\cal L}^{x,z}+{\cal L}^{z,x}+{\cal L}^{y,z}+{\cal L}^{z,y}\right)\label{eq:TFIM_noise}
\end{equation}
when $\gamma$ exceeds the critical value $\gamma_c=1$. Going beyond these particular choices of Hamiltonian and Lindbladian, a companion problem is to find the optimal gauge that cancels as many terms as possible for fixed Hamiltonian and noise terms. Let $\boldsymbol{\lambda}$ be the set of real numbers that linearly parameterize the gauge transform $\Lambda$.
Our goal is to determine the optimal $\boldsymbol{\lambda}$ that minimizes the estimate for particle growth \eqref{eq_mu_prediction}, $\mu(\boldsymbol{\lambda}) \sim \sum_{C \neq C'} \max (0, - [M + \Lambda(\boldsymbol{\lambda})]_{CC'})$.
Computing the $\max$ function can be cast as a linear minimization procedure over slack variables $t_{CC'} \in \mathbb{R}$ constrained by $t_{CC'} \geqslant 0$ and $t_{CC'} \geqslant -[M + \Lambda(\boldsymbol{\lambda})]_{CC'}$.
The augmented minimization cost function becomes $\mu(\boldsymbol{\lambda}, \boldsymbol{t}) = \sum_{C\neq C'} t_{CC'}$, together with the previous constraints.
Since both the cost function and the constraints are linear in $\boldsymbol{\lambda}$ and $\boldsymbol{t}$, the minimization corresponds to a linear programming problem that can be solved efficiently exactly \footnote{In our numerical simulations, we always perform this minimization procedure before the runs begin.
Indeed, finding the optimal gauge is generally necessary to observe the correct transition to classicality.}.

In Fig.~\ref{fig_tfim} we show numerical simulations of the TFIM in 2D for a global quench ($J = 1/2, h = 1$).
The initial state consists of all-to-all entangled Bell pairs:
\begin{equation}\label{eq_bell_pairs_state_main}
  \ket{\psi_0} = \bigotimes_{k = 1}^{N/2} \frac{1}{\sqrt{2}} 
  \left( \ket{+z,+z} 
  + e^{i\pi/4} \ket{-z,-z}
  \right)_{(j_k,l_k)} 
  \ ,
\end{equation}
where $\{(j_k, l_k)\}_{k=1}^{N/2}$ is a set of disjoint site pairs.
With non-local pairings, the initial state is volume-law entangled and highly magic~\cite{chitambar2019quantum,liu2022manybody,Leone_2022,Leone_2024}.
The noise term is given by \eqref{eq:TFIM_noise}.
We see that above a critical value $\gamma_{c} = 1$, the system transitions
to a phase of classical computational complexity.
In Fig.~\ref{fig_tfim}(a), we observe that the rate of particle growth is quantitatively close to the heuristic prediction of \eqref{eq_mu_prediction}, becoming zero for $\gamma > \gamma_c$.
In Fig.~\ref{fig_tfim}(b), we display perfect numerical agreement between the CTMC simulations and exact diagonalization (ED) simulations for both short- and long-range two-point correlations.
We also show that the classical phase allows the simulation of systems with thousands of qubits, far beyond what is accessible with ED.

\prlsection{Conclusion}
We have developed a stochastic representation of quantum dynamics in terms of negative Markov chains, establishing an exact mapping between unitary and open quantum evolution and a classical process on an enlarged configuration space. Within this framework, the computational complexity of quantum dynamics is directly tied to the proliferation of signed stochastic particles that encode the full quantum dynamics.

Our main result shows that, for a broad class of local and pairwise interacting systems, the addition of finite noise, combined with an appropriate gauge choice, can render all transition rates positive above a critical threshold. This induces a sharp transition to a regime described by a genuine classical CTMC, enabling efficient Monte Carlo simulations. Notably, this transition is governed purely by the structure of the stochastic representation and is not directly linked to standard complexity indicators such as entanglement.

Beyond providing a constructive route to the classical simulation of noisy quantum systems, our work identifies a new mechanism for the emergence of classicality in open quantum dynamics. Promising directions include the development of controlled approximate schemes---e.g., projections onto restricted configuration spaces in the spirit of tensor-network methods \cite{L_White1992,M_Orus2014}---and adaptive strategies to mitigate replica proliferation, such as resampling techniques from classical Monte Carlo \cite{N_Doucet2001,DelMoral2013_MonteCarlo}. These extensions offer a promising route toward bridging the gap between exact classical simulability and efficient approximate methods in the pre-threshold regime.

\prlsection{Acknowledgments}
JDN and HL are funded by the ERC Starting Grant 101042293 (HEPIQ) and the ANR-22-CPJ1-0021-01. TJ is funded by the ANR-25-CE57-2088 JCJC (QuDi). T.J. thanks Noah Orazi and Raphaël Chetrite for insightful discussions and collaborations on related subjects.

\bibliography{apssamp_sarang,biblio_Tony}

\clearpage

\appendix
\onecolumngrid

\section{Single-Qubit example}

\begin{figure}[ht]
\centering
\includegraphics[width=0.65\textwidth]{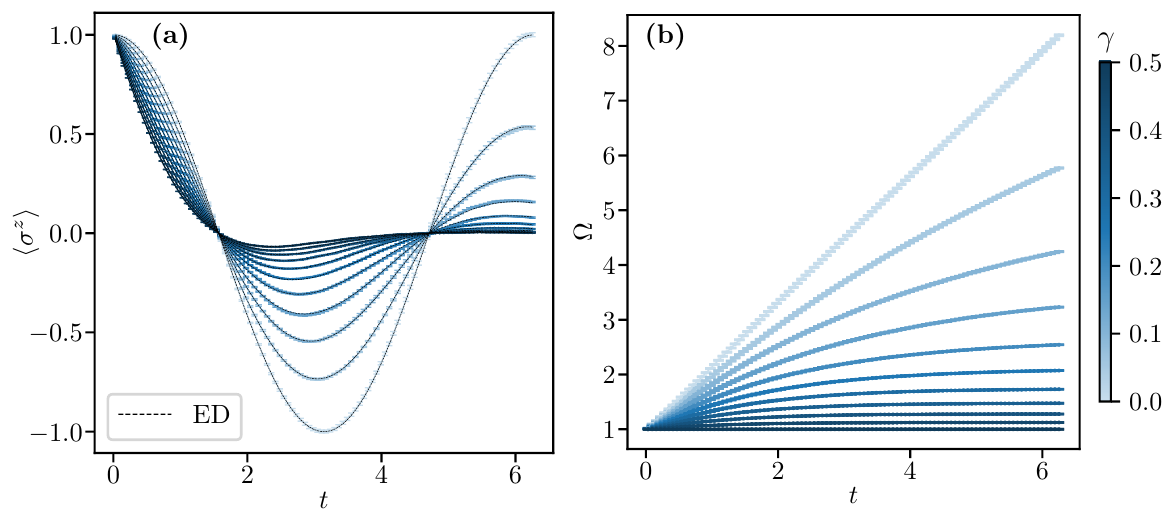}
\caption{
(a) Time-evolution of $\left\langle \sigma_{z}(t)\right\rangle $ for
the Lindblad evolution (\ref{eq:Lindbladsinglesite}) with initial
state $\left|\psi\left(t=0\right)\right\rangle =\left|+z\right\rangle $ and $\tau = 1/2$.
We compare the exact solution with the one obtained from the simulation
of the (negative) Markov process defined in (\ref{eq:Markovtransitionmatrix_singlesite})
and see perfect agreement. 
(b) Growth of the total particle number as a function of time. 
We see that above the critical value for which $\gamma=\tau$, all the transition weights
become positive and thus, the problem is entirely classical.}
\label{fig:single_spin}
\end{figure}

In this appendix, we illustrate the transition to classical CTMC on
the simplest example of a single-qubit undergoing unitary rotation
and dephasing: 
\begin{equation}
d_{t}\rho=-i\tau\left[\sigma_{x},\rho\right]+\gamma\left(\sigma_{x}\rho\sigma_{x}-\rho\right).\label{eq:Lindbladsinglesite}
\end{equation}
We consider for the initial state $\left|\psi\left(t=0\right)\right\rangle =\left|+z\right\rangle $
and $\gamma>0$. Since we can discard the $x$ degrees of freedom
for this example, the classical configuration space comprises here
the 4 states $\left\{ +,-\right\} \times\left\{ y,z\right\} $. The
Markov transition matrix restricted to this space, reads (the states are arranged in the order $(+y),(-y),(+z),(-z)$)
\begin{equation}
M=\begin{pmatrix}-\gamma & \gamma & \tau & -\tau\\
\gamma & -\gamma & -\tau & \tau\\
-\tau & \tau & -\gamma & \gamma\\
\tau & -\tau & \gamma & -\gamma
\end{pmatrix}.
\end{equation}
For $\gamma=0$, this unitary dynamics leads to oscillation of the
probability $p_{\pm,z}\left(t\right)=\frac{1}{4}\left(1\pm{\rm cos}(2\tau t)\right)$
and to, linear in $t$, unbounded growth of the absolute number of
replicas, see Fig.~\ref{fig:single_spin}. For $\gamma$ finite,
we see that the contribution from the pure jump term in the Lindblad
evolution leads indeed to only positive terms on the off-diagonal
but do not cancel explicitly the negative terms elsewhere. However,
let $\Lambda$ be the gauge transformation 
\begin{equation}
\Lambda=\begin{pmatrix}-1 & -1 & 1 & 1\\
-1 & -1 & 1 & 1\\
1 & 1 & -1 & -1\\
1 & 1 & -1 & -1
\end{pmatrix}.
\end{equation}
One can explicitly check that the sum over the rows of each column
is $0$ and that $\forall C$ $\sum_{C'}\Lambda_{CC'}\mathbb{P}_{C'}=0$,
making it a good gauge transformation according to our criteria from
the main text. Consequently, adding this gauge to $M$ does not change
the averaged process but does change the transition rules: 
\begin{equation}
M+\gamma\Lambda=\begin{pmatrix}-2\gamma & 0 & \gamma+\tau & \gamma-\tau\\
0 & -2\gamma & \gamma-\tau & \gamma+\tau\\
\gamma-\tau & \gamma+\tau & -2\gamma & 0\\
\gamma+\tau & \gamma-\tau & 0 & -2\gamma
\end{pmatrix}\label{eq:Markovtransitionmatrix_singlesite}
\end{equation}
and we see that for $\gamma>\left|\tau\right|$, all the terms apart
from those on the diagonal become positive. In this regime, the system
is described by a CTMC with only positive weights so no replica is
required to evolve the system. We obtain exactly the CTMC description
of a \emph{classical system}. On Fig.~\ref{fig:single_spin} we show
the evolution of the system described by the transition matrix Eq.~(\ref{eq:Markovtransitionmatrix_singlesite})
along with the particle production with and without the gauge transform
as a function of $\gamma$. We see that for $\gamma>\left|\tau\right|$
the number of particle doesn't grow at all making the system entirely
classical.

\section{Minimization procedure}

In this appendix, we provide more details on the minimization procedure
leading to the optimal noise that makes a given unitary dynamics generated
by $H$ classical. Optimal here means that the coefficients weighting
the noise term minimizes a given cost function. 

An important thing to notice is the overcounting of local terms in
the Markov transition matrix. Indeed, the latter acts on \emph{links}. Consequently, in dimension $d$, for cubical lattices, the local terms act $2d$ times on a given site. Hence, one needs to make a rescaling
of a factor $1/2d$ when going from local terms in the Lindblad generator
to the Markov matrix. The different steps are then as follow. Consider
an Hamiltonian of the form: 
\begin{equation}
H=\sum_{j,a}h_{j}^{a}\sigma_{a}^{j}+\sum_{j\neq k,a,b}J_{a,b}^{j,k}\sigma_{a}^{j}\sigma_{b}^{k}\label{eq:Hamiltonian-1}
\end{equation}
that we denote for simplicity 
\begin{equation}
H=hH_{{\rm loc}}+JH_{{\rm pair}}
\end{equation}
where $H^{{\rm loc}}$ comprises all possible local terms, $H^{{\rm pair}}$
all possible pairwise terms and $h$ and $J$ are the corresponding
weights of these terms. 

The associated Markov matrix is 
\begin{equation}
M=\frac{h}{2d}M_{{\rm loc}}^{H}+JM_{{\rm pair}}^{H}.
\end{equation}
Let $\nu{\cal L}_{{\rm loc}}:=\sum_{j,a}\nu_{j}^{a}{\cal L}^{*}\left[\sigma_{j}^{a}\right]$
and $\gamma{\cal L}_{{\rm pair}}:=\sum_{j,k,a,b}\gamma_{j,k}^{a,b}{\cal L}^{*}\left[\sigma_{j}^{a}\sigma_{k}^{b}\right]$.
Our goal is to find the coefficients $\nu$, $\gamma$$\geq0$ such
that 
\begin{equation}
\frac{h}{2d}M_{{\rm loc}}^{H}+JM_{{\rm pair}}^{H}+\frac{\nu}{2d}\left(\Lambda^{1}+M_{{\rm local}}^{{\cal L}}\right)+\gamma\left(\Lambda^{2}+M_{{\rm pair}}^{{\cal L}}\right)
\end{equation}
 has positive coefficients on the off-diagonal while minimizing a
cost function such as e.g. the $L^{1}$ norm
\begin{equation}
\sum_{j}\left|\nu_{j}^{a}\right|+\sum_{i,j,a,b}\left|\gamma_{j,k}^{a,b}\right|.
\end{equation}
This problem can be cast as a linear optimization problem that
is trivially solved numerically as the number of possible terms is
only 15. 

\section{Numerical details for the simulation of negative Markov chains.\label{App:Numerics}} 

Here, we present more details on the numerical implementation of the quantum CTMCs, based on Monte Carlo sampling of runs of agent/particle-based simulations.

\subsection{Simulating the dynamics}

For any given particle, one needs to simulate transitions into other configurations, given by the negative terms of \eqref{eq:probabilitesparticle/antiparticle}, corresponding to escape/outflux rates.
Apart from the creation rate $V_C$, the other positive terms in \eqref{eq:probabilitesparticle/antiparticle}, corresponding to particle influx, are implicitly handled by simulating the outflux events.
It then is useful to define 
\begin{equation}\label{eq_escape_rate_per_conf}
  \tau_{C} = \sum_{C' \neq C} (M^+_{C' C} + M^-_{C'C}) = \sum_{C' \neq C} |M_{C'C}| \ ,
\end{equation}
the total escape rate per configuration.
There are different strategies of implementing the Markov chain rules of  \eqref{eq:probabilitesparticle/antiparticle} with the proper rates.
Perhaps the simplest approach entails setting a discrete constant time step $\Delta t$ and looping through all particles at a time $t$, deciding to act on each one with probability $\Delta t\tau_C$ given the particle configuration $C$.
However, one needs to enforce that each time step registers at most one event on average, so as not to accumulate errors.
This means that $\Delta t$ should be adaptive to the time-dependent particle number.
Another approach that naturally solves this issue is to have precisely one event per time-step, but instead randomly sample $\Delta t$ in an adaptive way, known as the Gillespie algorithm \cite{I_Gillespie1977}. 
Overall, the algorithm can be implemented in the following steps:

\begin{enumerate}
  \item At a given time $t$, compute the total escape timescale, $\tau_{\mathrm{tot}} = \sum_{C}  \tau_C(n_C^\circ + n_C^\bullet)$.
  \item Draw a time step from the exponential distribution $\Delta t \sim \mathrm{Exp} (\tau_{\mathrm{tot}})$ and update $t \rightarrow t + \Delta t$.
  \item Choose a particle according to the  probability distribution $P_1(C, \bullet / \circ) = n_C^{\bullet / \circ}\tau_C/\tau_{\mathrm{tot}} $.
  \item After having chosen a particle in a given configuration $C$, choose a configuration for the destination $C'$ and the $M$ sign with probability distribution $P_2(C', \pm ) = M^{\pm}_{C'C} / \tau_C$.
    \begin{enumerate}
      \item If the sign $+$ was chosen, simply move the particle to configuration $C'$.
	To implement the annihilation rule, if the opposite particle already exists in $C'$, delete both.
      \item If the sign $-$ was chosen, create a particle in  configuration $C'$ and opposite type.
	To implement the annihilation rule, if a particle opposite to the created one already exists, delete both.
	Additionally, copy the original particle in configuration $C$ to handle the creation term $V_C$ with the proper rate.
    \end{enumerate}
\end{enumerate}

The efficiency of the algorithm will depend on the complexity of sampling $P_1$ and $P_2$, as well as particle lookup, creation, and deletion.
We found that binary search trees (BSTs) efficiently performed all particle operations.
In particular, the binary tree should be balanced and dynamic, so we used a \textit{treap} \cite{seidel1996} (it is possible that a simpler dynamic BST would be naturally balanced due to trajectory ergodicity).
The treap is sorted from left to right according to the flattened configuration label $C$.
Each node of the tree corresponds to a configuration $C$, where we also store $n_C^{\circ/\bullet}$ and $\tau_C$.
Nodes where $n_C^\circ = n_C^\bullet = 0$ are removed from the BST.
Additionally, we can efficiently store relevant subtree (the tree under and including any given node) information, such as subtree size (number of nodes), subtree number of particles, and subtree cumulative escape rate $\sum_{C \in \rm{subtree}} \tau_C(n_C^\circ + n_C^\bullet)$.
Overall, the size of the BST corresponds to the number of occupied particle configurations $\Omega_{\rm occ}$.
Consequently, all operations such as adding, moving, or removing particles have time complexity $\mathcal{O}(\log \Omega_{\rm occ})$.

The total escape rate $\tau_{\rm tot}$ is automatically tracked by the BST, so steps 1. and 2. are performed in $\mathcal{O}(1)$ time.
Assuming all $\tau_C$ are comparable, we get $\tau_{\rm tot} \sim \Omega$, therefore the mean time step is $\overline{\Delta t} \sim 1/\Omega$.
Step 3. of the Gillespie algorithm can be performed with a binary search using the subtree cumulative escape rates in $\mathcal{O}(\log \Omega_{\rm occ})$ time.

We also need to store the rate matrix $M$ and sample transitions $P_2$ of step 4.
Even though the matrices $M_{C'C}$ are sparse if the dynamics are few-body local, they still contain an exponentially larger amount of non-zero elements, so they cannot be globally stored in memory.
However, each contribution to $M$ comes from few-body interactions, acting on (usually) $\mathcal{O}(N)$ groups of few interacting sites.
For each configuration $C$ (each node of the particle BST), all possible allowed transitions could be stored in another BST with $\mathcal{O}(N)$ nodes, sorted according to the spatial structure of the system.
Transitions could then be sampled according to $P_2$ in $\mathcal{O}(\log N)$ time.
This comes with an added memory cost of having to store an internal rate BST in each node of the particle BST.
Otherwise and more simply, we can just store all the unique few-body contributions to $M$ and loop through those for each set of interacting sites to sample $P_2$ in $\mathcal{O}(N)$ time with no more memory requirements.

Taking all the steps of the algorithm into account, to simulate a finite time interval $T = \mathcal{O}(1)$, it takes at least ${\cal O}\left(\Omega\log\Omega_{\rm occ}\log N\right)$ time with our current choice of architecture.

\subsection{Sampling the initial state}

We have understood how to time-evolve particles according to \eqref{eq:probabilitesparticle/antiparticle}, but we also need to know how to initialize them at $t = 0$. 
Essentially, for any given initial state $\rho_0$, we need to be able to efficiently sample from the initial probability distribution  
$p_C(0) = {\rm tr} (\rho_0 \mathbb{P}_{C}) / 3^N$. 
Each run starts with a single initial particle.
Since $p_C \geqslant 0$, initially we need only sample particles of type $\bullet$.
We will explicitly address this procedure for two types of initial states.

Let us suppose the initial state in the evolution is one of the configuration vectors $\ket{C_0} = \bigotimes^N_{i=1} \ket{s_i^0 \beta_i^0}$, where $s_i \in \{+,-\}$ and $\beta_i \in \{x,y,z\}$.
The initial probability distribution is then given by
\begin{equation}\label{eq_initial_probabilities}
  \begin{aligned}
    p_C(0) & = \frac{1}{3^N} {\rm tr}[\mathbb{P}_{C_0} \mathbb{P}_{C}] 
    = \frac{1}{3^N} \prod_i | \langle s_i^0 \beta_i^0 | s_i \beta_i \rangle |^2 
     = \frac{1}{3^N} \prod_i \left[\delta_{s_i^0 s_i} \delta_{\beta_i^0\beta_i} + \frac{1}{2}( 1 - \delta_{\beta_i^0 \beta}) \right] \ ,
  \end{aligned}
\end{equation}
There is an exponentially large number in $N$ of non-zero initial probabilities.
Still, a simple way of sampling particles with the correct probability distribution \eqref{eq_initial_probabilities} is to loop over each site $i \in \{1, \dots N\}$ and 
\begin{enumerate}
  \item with probability $1/3$, choose the orientation $(s_i, \beta_i) = (s_i^0, \beta_i^0)$.
  \item else choose a random sign $s_i$ and a random axis $\beta_i \neq \beta_i^0$.
\end{enumerate}
The chosen particle configuration is $C = (s_1, \beta_1, \dots, s_N, \beta_N)$.

We can also choose initial states that are volume-law entangled and have a high degree of quantum magic.
The time evolution of these states would be hard to simulate with both tensor networks and Pauli propagation methods.
One possibility is to consider non-local phase-shifted Bell pairs.
If we define a set of disjoint site pairs $\{(j_k, l_k)\}_{k=1}^{N/2}$ and set of associated phases $\{\theta_k\}_{k=1}^{N/2}$, the state is expressed as
\begin{equation}\label{eq_bell_pairs_state}
  \ket{\psi_0} = \bigotimes_{k = 1}^{N/2} \frac{1}{\sqrt{2}} 
  \left( \ket{+z}_{j_k} \ket{+z}_{l_k} 
  + e^{i\theta_k} \ket{-z}_{j_k} \ket{-z}_{l_k}\right) 
  \ .
\end{equation}
The probability distribution can then be expressed as 
\begin{equation}
  \begin{aligned}
    p_C(0) & = \frac{1}{3^N} \bra{\psi_0} \mathbb{P}_{C} \ket{\psi_0} \\
    & = \frac{1}{3^N 2^{N/2}} \prod_{k = 1}^{N/2} \Big[ 
    |\langle +z{+}z | s_{j_k} \beta_{j_k} s_{l_k} \beta_{l_k} \rangle|^2  + 
    |\langle -z{-}z | s_{j_k} \beta_{j_k} s_{l_k} \beta_{l_k} \rangle|^2 
    \\
    & \hspace{2.5cm} + 
    2 \cos\theta_k  \left( \langle +z {+}z | \mathbb{P}^{s_{j_k} \beta_{j_k}}\mathbb{P}^{s_{l_k} \beta_{l_k}} |-z {-}z\rangle \right) 
    -2 \sin\theta_k \mathrm{Im} \left( \langle +z {+}z | \mathbb{P}^{s_{j_k} \beta_{j_k}}\mathbb{P}^{s_{l_k} \beta_{l_k}} |-z {-}z\rangle \right) 
    \Big]\\
    & = \frac{1}{36^{N/2}} \prod_{k = 1}^{N/2} 
    \Big[ 1 + 
    s_{j_k} s_{l_k} \cos \theta_k \left( 
    \delta_{\beta_{j_k} x} \delta_{\beta_{l_k} x} -
    \delta_{\beta_{j_k} y} \delta_{\beta_{l_k} y} 
    \right) \\ 
    & \hspace{2.4cm} + 
    s_{j_k} s_{l_k} \sin \theta_k \left( 
    \delta_{\beta_{j_k} x} \delta_{\beta_{l_k} y} +
    \delta_{\beta_{j_k} y} \delta_{\beta_{l_k} x} 
    \right) 
    + \delta_{\beta_{j_k} z} \delta_{\beta_{l_k} z} 
    ( 2 \delta_{s_{j_k} s_{l_k} } - 1)
    \Big]
  \end{aligned}
\end{equation}
We can sample the distribution with the following random choice of spin orientations for each pair with label $k$.
\begin{center}
    \begin{tabular}{|c|c|}
         \hline
         Probability & Set for random choice of $(s_{j_k} \beta_{j_k}, s_{l_k} \beta_{l_k})$ \\
         \hline
         $(1 + \cos\theta_k)/9$ & $\{(+x, +x), (-x,-x), (+y,-y), (-y,+y) \}$ \\
         \hline
         $(1 - \cos\theta_k)/9$ & $
         \{ (+x, -x), (-x,+x), (+y,+y), (-y,-y) \}
         $ \\
         \hline
         $(1 + \sin\theta_k)/9$ & $
         \{ (+x, +y), (+y,+x), (-x,-y), (-y,-x) \}
         $ \\
         \hline
         $(1 - \sin\theta_k)/9$ & $
         \{ (+x, -y), (-y,+x), (-x,+y), (+y,-x) \}
         $ \\
         \hline
         $1/9$ & $
         \{ (+z, +z), (-z,-z)\}
         $ \\
         \hline
         $4/9$ & $
         (\beta_{j_k}, \beta_{l_k}) \in  \{(x, z), (z,x), (y,z), (z,y)\}
         $, $s_{j_k},s_{l_k} \in \{+,-\}$ \\
         \hline
    \end{tabular}
\end{center}
If there is an unpaired site, we can fix its orientation and sample according to our first initial state example.
The chosen particle configuration is $C = (s_1, \beta_1, \dots, s_N, \beta_N)$.

After finding an initial configuration $C$, we can compute the escape rate $\tau_C$.
We might come across stationary particles with $\tau_C = 0$.
In that case, we do not explicitly perform the time evolution but take into account their contribution to $p_C$, before continuing to the next run.

\subsection{Computing observables}

The final ingredient for implementing efficient simulations is computing expectation values of observables.
To do so, we expand the observable in terms of projectors $\hat{O} = \sum_C O_C \mathbb{P}_C$ (the expansion is not unique).
The expectation value then becomes
\begin{equation}
  {\rm tr}[\rho_t \hat{O}] = 3^N \sum_C O_C p_C = 3^N \sum_C O_C \mathbb{E}[n^\bullet_C(t) - n_C^\circ(t)] \ .
\end{equation}
Note that this sum is over an exponentially large number of configurations, so one should loop through each active particle instead and add their contribution.
We can efficiently keep track of the observable value at all times if we update with the contribution of each particle that gets added, moved, or destroyed.
The complexity of updating the observable tracker at each time step corresponds to that of computing $O_C$ (which cannot be precomputed due to the exponential number of configurations).

We consider specifically expectation values of Pauli operators, 
\begin{equation}\label{eq_pauli_expectation}
  \begin{aligned}
    {\rm tr}[\rho_t \sigma^i_{\alpha_i}] & = \frac{1}{3^{N-1}} {\rm tr}\left[\rho_t 
    \prod_{j < i}\left(\sum_{s_j \beta_j} \mathbb{P}_j^{s_j \beta_j}\right)
    \left(\mathbb{P}_i^{+ \alpha_i} - \mathbb{P}_i^{- \alpha_i}\right)
    \prod_{j > i}\left(\sum_{s_j \beta_j} \mathbb{P}_j^{s_j \beta_j}\right)
    \right] \\
    & = 3 \sum_{(s_1, \beta_1, \dots, s_{N}, \beta_{N})} 
    s_i \delta_{\alpha_i \beta_i}
    \mathbb{E}[n^\bullet_C(t) - n_C^\circ(t)] \ ,
  \end{aligned}
\end{equation}
with $C = (s_1, \beta_1, \dots, s_N, \beta_N)$ and $\mathbb{P}_i^{s_i \alpha_i} = (\mathds{1} + \sigma^i_{\alpha_i})/2$.
Since for a given particle of fixed $C$, its contribution corresponds simply to $3s_i \delta_{\alpha_i \beta_i}$, the observable expectation value tracker can be updated at each time step in $\mathcal{O}(1)$ time.
Note that we chose the expansion of the Pauli operator that was more uniformly distributed in the configuration space, since this should minimize the amount of sampling needed to compute the expectation value.
We can generalize this formulation for expectation values of Pauli strings, such as 
\begin{equation}
  \begin{aligned}
    {\rm tr}\left[ \rho_t \prod_{j \in A}\sigma^j_{\alpha_j} \right] & 
    = \frac{1}{3^{N-N_A}} {\rm tr}\left[\rho_t 
    \prod_{j \notin A}\left(\sum_{s_j \beta_j} \mathbb{P}_j^{s_j \beta_j}\right)
    \prod_{j \in A} \left(\mathbb{P}_j^{+ \alpha_j} - \mathbb{P}_j^{- \alpha_j}\right)
    \right] \\
    & = 3^{N_A} \sum_{(s_1, \beta_1, \dots, s_{N}, \beta_{N})} 
    \left[
      \left( \prod_{j \in A} s_j  \delta_{\alpha_j \beta_j} \right)
      \mathbb{E}[n^\bullet_C(t) - n_C^\circ(t)] 
      \right]
    \ ,
  \end{aligned}
\end{equation}
where the Pauli string is not trivial in a region $A$ of length $N_A$.
The observable expectation value tracker can be updated in each time step in $\mathcal{O}(N_A)$ time.

\section{Proof of the main theorem and corollary}\label{sec:Proof}

In this appendix, we provide more details on the proof of the main
theorem and corollary. The idea of the proof is to find an appropriate
gauge transformation such that negative weights from the Hamiltonian
are entirely compensated by the positive weights coming from the open
system terms. Since the elements of the Markov matrix of the latter
belong to $\mathfrak{L}$, our goal is to find a gauge transform that
brings the negative weights of the Hamiltonian part $H$ in the form
of the terms in $\mathfrak{L}$.

For single-spin terms, consider the gauge transform 
\begin{equation}
\Lambda^{a}\left(\mathbb{P}_{s,\alpha}\right)=\varepsilon_{a,\alpha,\alpha'}^{2}\mathbb{I}_{\alpha'}-\bar{\delta}_{a,\alpha}\mathbb{I}_{\alpha}
\end{equation}
$\mathbb{I}_\alpha$ is the identity operator and the subscript is there to emphasize that we use here the decomposition $\mathbb{I}_\alpha=\mathbb{P_{+,\alpha}-\mathbb{P_{-,\alpha}}}.$ The dynamics with the gauge is then expressed as
\begin{align}
\left(\tau{\rm ad}_{\sigma_{a}}+\gamma\left(\Lambda^{a}+{\cal L}_{a}^{*}\right)\right)\left[\mathbb{P}_{s,\alpha}\right] & =\varepsilon_{a,\alpha,\alpha'}^{2}\left(\gamma\mathbb{I}_{\alpha'}-s\tau\varepsilon_{a,\alpha,\alpha'}\sigma_{\alpha'}\right)-\bar{\delta}_{a,\alpha}\gamma\left(\mathbb{I}_{\alpha}+s\sigma_{\alpha}\right)
\end{align}
and we see that for $\gamma\geq\tau$, all the off-diagonal terms
become positive. We now turn to the pairwise interactions. Recall
the action of the Hamiltonian $H=J\sigma_{a}\sigma_{b}$ on a set
of projectors: 
\begin{align}
-2{\rm ad}_{\sigma_{a}^{j}\sigma_{b}^{k}}\left(\mathbb{P}_{s_{1},\alpha}^{j}\mathbb{P}_{s_{2},\beta}^{k}\right)= & s_{1}\varepsilon_{a,\alpha,\alpha'}\sigma_{\alpha'}^{j}\left(\sigma_{k}^{b}+s_{2}\delta_{b,\beta}\mathbb{I}\right)+j\leftrightarrow k.\label{eq:Hamiltonian_2sites-2}
\end{align}
in the main text. It is enough to only consider the first half of
this expression as the other one is just a permutation of indices.
Let
\begin{align}
2h^{a,b}\left[\mathbb{P}_{s_{1},\alpha}\mathbb{P}_{s_{2},\beta}\right]:= & -s_{1}\varepsilon_{a,\alpha,\alpha'}\sigma_{\alpha'}\left(\sigma_{b}+s_{2}\delta_{b,\beta}\mathbb{I}\right).\label{eq:Hamiltonian_2sites-1}
\end{align}
After a careful inspection, we see that an appropriate gauge transformation
is
\begin{align}
\Lambda^{a,b}\left[\mathbb{P}_{s_{1},\alpha}\mathbb{P}_{s_{2},\beta}\right] & :=\frac{1}{2}\left(\left(\varepsilon_{a,\alpha,\alpha'}^{2}\mathbb{I}_{\alpha'}-\bar{\delta}_{a,\alpha}\mathbb{I}_{\alpha}\right)\left(\mathbb{I}_{b}+\delta_{b,\beta}s_{2}\sigma_{b}\right)+\bar{\delta}_{a,\alpha}\mathbb{I}_{\alpha}\left(\delta_{b,\beta}\left(\mathbb{I}_{{\rm tot}}-3\mathbb{I}_{b}\right)+\mathbb{I}_{b}-\mathbb{I}_{\beta}\right)\right)
\end{align}
with $\mathbb{I}_{{\rm tot}}:=\mathbb{I}_{x}+\mathbb{I}_{y}+\mathbb{I}_{z}$.
One can explicitly check that $\Lambda$ is indeed a valid gauge transform,
i.e. $\sum_{C}\Lambda_{CC'}^{a,b}\mathbb{P}_{C}=0$ $\forall C'$
and $\sum_{C'}\Lambda_{CC'}^{a,b}=0$ $\forall C$. Combining these terms, one arrives at 
\begin{align}
\left(Jh^{a,b}+\gamma\Lambda^{a,b}\right)\left[\mathbb{P}_{s_{1},\alpha}\mathbb{P}_{s_{2},\beta}\right]= & \frac{1}{2}\Big(\left(\gamma\varepsilon_{a,\alpha,\alpha'}^{2}\mathbb{I}_{\alpha'}-Js_{1}\varepsilon_{a,\alpha,\alpha'}\sigma_{\alpha'}\sigma_{b}\right)\left(\mathbb{I}_{b}+\delta_{b,\beta}s_{2}\sigma_{b}\right)\\
 & +\gamma\bar{\delta}_{a,\alpha}\mathbb{I}_{\alpha}\left(\delta_{b,\beta}\mathbb{I}_{{\rm tot}}-\left(\delta_{b,\beta}\left(3\mathbb{I}_{b}+s_{2}\sigma_{b}\right)+\mathbb{I}_{\beta}\right)\right)\Big).\nonumber 
\end{align}
The first term in the right-hand side of the previous expression is always positive for $\gamma\geq\left|J\right|$.
The negative contributions are then entirely contained in the second term, and all belong to the set $\mathfrak{L}$.
Since all the possible terms in $\mathfrak{L}$ can be obtained from the set of Lindblad jump operators that we consider, this proves our theorem.

The proof for the corollary comes from the fact that the uniform two-site
noise $\gamma\left(\sum_{a}{\cal L}^{a*}+\sum_{a,b}{\cal L}^{a,b*}\right)$
comprises all the terms in $\mathfrak{L}$. Thus, there is necessarily a critical value $\gamma_{c}$ above which the weights all become
positive.
\begin{figure}
\centering{}\includegraphics[width=1\textwidth]{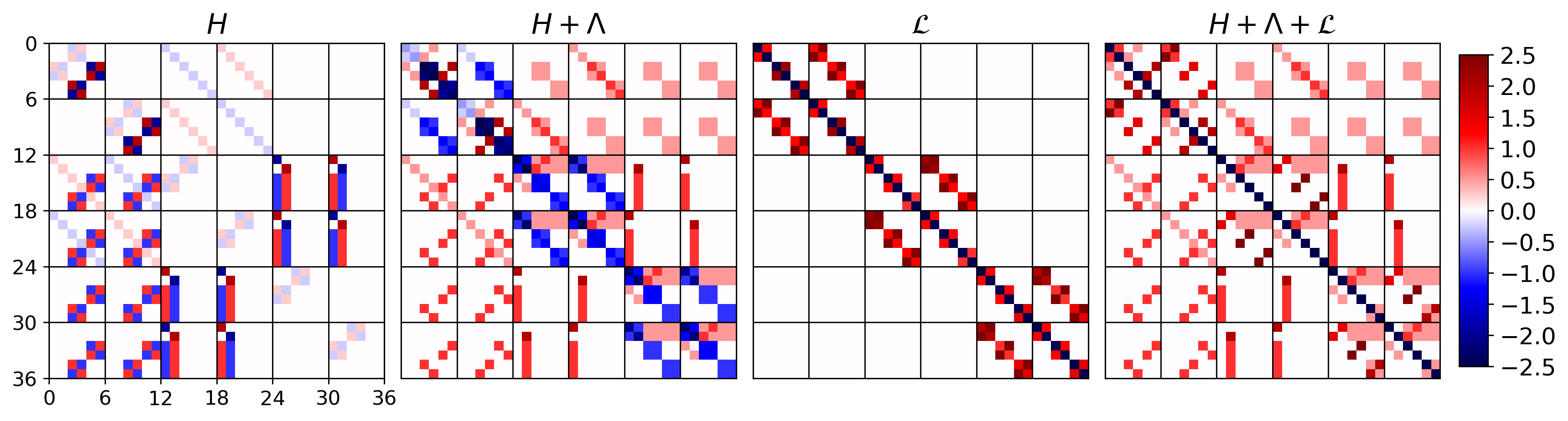}\caption{Illustration of the procedure rendering the Markov transition matrix
classical for the example of the TFIM (\ref{eq:TFIM_Hamiltonian})
with noise terms (\ref{eq:TFIM_noise}). The rows and the columns
are indexed as $\left(\left\{ x,y,z\right\} \times\left\{ +,-\right\} \right)^{2}$ (note that the indexing convention is reversed compared to the main text).
From left to right: 1) The bare transition matrix of $H$ with negative
weights outside of the diagonal that eventually leads to the growth
of replicas. 2) Gauge transform $H+\Lambda$ casting the negative
terms in the form of $\mathfrak{L}$. 3) Markov matrix corresponding
to the Lindblad term ${\cal L}$. 4) Combining all terms together
we get a Markov transition matrix with entirely positive weights outside
of the diagonal. }\label{fig:Classicalization}
\end{figure}
To end this section, we show graphically on the explicit example of
the TFIM, Eq.~(\ref{eq:TFIM_Hamiltonian}) of the main text, the
different transformations applied on the Markov matrix to make it
classical, see Fig.~\ref{fig:Classicalization}.

\section{Heuristic analytical predictions for particle growth and saturation}\label{App:Predictions}

In this appendix, we present analytical derivations for particle number growth rates and saturation values.
We will impose the heuristic assumptions that single-particle trajectories are ergodic and that, at sufficiently late times (post-thermalization), the probability of finding a particle in any given configuration is uniform.
In that case, the rate of particle creation is the average over the columns of $M$ of the absolute values of the negative off-diagonal terms 
\begin{equation}
  \tau_{\rm cr} = \frac{2Nd}{6^{k}}\sum_{j\neq j'}M_{jj'}^{\left(k\right)-}  
  \ .
\end{equation}
The prefactor assumes that there are $Nd$ identical local rate matrices $M^{(k)}$ contributing to the global $M$, each acting on $k$ sites (as is the case for $k$-body local TI models).
The factor of $2$ comes from the fact that each negative event generates two particles.

Annihilations occur in pairs in each transition if the destination configuration was occupied by a particle of opposite type.
Therefore, the rate of annihilation corresponds to the product of the average escape rate and the probability of finding at least one particle of opposite type in the destination configuration
Assuming uniform particle distribution, that probability corresponds to $1 - (1 - 6^{-N}/2)^\Omega$.
The overall annihilation rate becomes
\begin{equation}
  \tau_{\rm an} = 2  \frac{Nd}{6^k}
  \sum_{j \neq j'} \left|M^{(k)}_{jj'} \right|
  \left[ 1 - \left(1 - \frac{6^{-N}}{2} \right)^{\Omega}\right] 
  \ .
\end{equation}

When the number of particles saturates, the annihilation and creation rates must be equal, resulting in a condition for the steady-state particle number.

\begin{equation}
  \begin{aligned}
    \tau_{\rm cr} = \tau_{\rm an} 
    & \Leftrightarrow
    \left(1 - \frac{6^{-N}}{2} \right)^\Omega =  
    1 - 
    \frac{\sum_{j \neq j'} 
    M^{(k)-}_{jj'} }{
      \sum_{j \neq j'} \left|M^{(k)}_{jj'} \right|}
    \\
    & \Leftrightarrow
    \Omega = 
    \log \left( 
    1 - 
    \frac{\sum_{j \neq j'} M^{(k)-}_{jj'} }{
      \sum_{j \neq j'} \left|M^{(k)}_{jj'} \right|}
    \right) 
    /\log\left(1 - \frac{6^{-N}}{2} \right)
    \\
    & \Leftrightarrow
    \Omega \approx - 2 (6^N)
    \log \left( 
    1 - 
    \frac{\sum_{j \neq j'} M^{(k)-}_{jj'} }{
      \sum_{j \neq j'} \left|M^{(k)}_{jj'} \right|}
    \right) 
    \equiv \Omega_{\rm sat}
    \\
  \end{aligned}
\end{equation}

\begin{figure}
    \centering
    \includegraphics[width=0.5\linewidth]{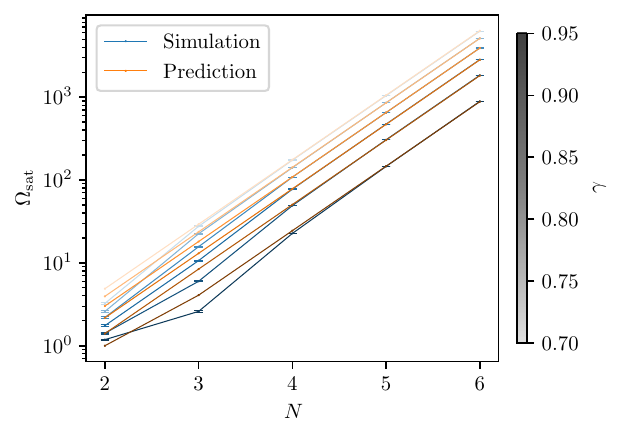}
    \caption{
    Saturation particle number for the TFIM ($J = 1/2, h = 1$) in 1D with noise given by \eqref{eq:TFIM_noise} with a prefactor $\gamma$ (classical at $\gamma \geq 1$). 
    The initial state is given by $\ket{+x}^{\otimes N}$.
    }
    \label{fig:pnum_sat}
\end{figure}

As seen in Fig.~\ref{fig:pnum_sat}, even though some assumptions are not rigorously justified, this prediction is in good agreement with the numerics.
Before particle saturation, with $\Omega \ll \Omega_{\rm sat}$, one gets that $\tau_{\rm cr} \gg \tau_{\rm an}$, leading to \eqref{eq_mu_prediction}, with $\mu = \tau_{\rm cr}/N$.

\end{document}